\documentclass[twocolumn,prl,showpacs,preprintnumbers,amsmath,amssymb]{revtex4}

\usepackage{graphicx}
\usepackage{dcolumn}
\usepackage{bm}
\usepackage{amsmath}
\usepackage{subfigure}

\begin{document}

\title{Multiply periodic states and isolated skyrmions in an anisotropic frustrated magnet}

\author{A. O. Leonov}
\thanks
{Corresponding author} 
\email{a.leonov@rug.nl}
\author{M. Mostovoy}
\affiliation{Zernike Institute for Advanced Materials, University of Groningen, Nijenborgh 4,9747 AG Groningen,  The Netherlands}

\date{\today}

\begin{abstract}
{

Multiply periodic states appear in a wide variety of physical contexts, such as the Rayleigh-B\'enard convection, Faraday waves, liquid crystals, domain patterns in ferromagnetic films and skyrmion crystals recently observed in chiral magnets. 
Here we study a simple model of an anisotropic frustrated magnet and show that its zero-temperature phase diagram contains numerous multi-$q$ states including the skyrmion crystal. 
We clarify the mechanism for stabilization of these states, discuss their multiferroic properties and formulate rules for finding new skyrmion materials. 
In addition to skyrmion crystal, we find stable isolated skyrmions with topological charge 1 and 2.
Physics of isolated skyrmions in frustrated magnets is very rich. 
Their statical and dynamical properties  are strongly affected by the new zero mode - skyrmion helicity.

}
\end{abstract}


         
\maketitle

%


\section*{Introduction}

Magnetic skyrmion is a topological defect with a complex  non-coplanar spin structure \cite{Nagaosa2013}. Predicted more than 20 years ago \cite{Bogdanov1989}, skyrmions have been recently observed in conducting and insulating helimagnets under an applied magnetic field \cite{Muelbauer2009,Yu2010}. Magnetic skyrmion provides a physical realization of the idea that quantization of physical observables, such as electric and baryon charge, is rooted in topology \cite{Dirac1931,Skyrme1961}. The skyrmion spin texture induces one flux quantum of an effective magnetic field acting on spin-polarized electrons and magnons, which gives rise to topological Hall effects in charge and heat transport and at the same sets skyrmions into motion \cite{Lee2009,Neubauer2009,Zang2011,Schulz2012,Vanhoogdalem2013,Mochizuki2014}.  Low critical currents needed to manipulate skyrmions opened a new active field of research - skyrmionics, which has a goal of developing skyrmion-based  magnetic memory and data processing devices  \cite{Fert2013,Iwasaki2013,Zhou2014,Tomasello2014}. 

So far, skyrmions have only been found in a handful of materials, all with the non-centrosymmetric cubic B20 structure, in which the relativistic Dzyaloshinskii-Moriya interaction turns a collinear spin state into a helical spiral. Under an applied magnetic field the spiral transforms into the skyrmion crystal (SkX) with a triangular magnetic superlattice. Further increase of the magnetic field results in a transition to the saturated ferromagnetic state, in which isolated   skyrmions exist as stable topological defects \cite{Butenko10,Leonov11}.  

Skyrmions are close relatives of magnetic bubbles (cylindrical domains) and the transformation of the helical spiral into SkX is analogous to the field-induced transition between the stripe domain state and the bubble array taking place in thin ferromagnetic films. In the framework of Landau theory the bubble array is  described by three coexisting spin modulations with the wave vectors, $\bm q_1, \bm q_2$ and $\bm q_3$, which add to zero. This $3q$-state is stabilized by a non-linear interaction between these modulations and the uniform magnetization induced by an applied magnetic field \cite{Garel1982}.  In bulk chiral magnets SkX competes with the conical spiral state and an additional order-from-disorder mechanism - stabilization by thermal spin fluctuations - was invoked to explain why skyrmions are only  observed at elevated temperatures \cite{Muelbauer2009,Buhrandt2013}. 

Practical applications of skyrmions crucially depend on finding new classes of skyrmion materials and new microscopic mechanisms for their stabilization. Recently, Okubo {\it et al.} studied numerically an isotropic Heisenberg spin model on a (centrosymmetric) triangular lattice with competing spin interactions \cite{Okubo2012}. It shows a spiral ground state, which above a critical magnetic field transforms into the SkX. In addition, a state with two coexisting spirals was found in applied magnetic fields. These multi-$q$ states are induced by thermal fluctuations and appear at nonzero temperatures.

Here we discuss a new mechanism which lends stability to multiply periodic states even at zero temperature.
We show that an uniaxial magnetic anisotropy strongly affects spin ordering in the frustrated triangular magnet: Tuning magnetic field and anisotropy at zero temperature, we find eight different phases, five of which are multi-\textit{q} states. A large part of the phase diagram is occupied by the SkX.

We clarify the nature of the phases found in numerical simulations and present results of analytical studies of  instability of spiral states towards additional periodic modulations. We obtain the spectrum of low-lying excitations in all ground states, which can be measured  in magnetic resonance experiments, discuss their multiferroic properties and the possibility to observe large magneto-electric responses in multi-$q$ states.   

We also explore very rich physics of isolated skyrmions in frustrated magnets. We find that their monopole and toroidal moment densities oscillate with the distance from the skyrmion centre, which gives rise to alternation of attraction and repulsion between skyrmions, clustering of skyrmions in high magnetic fields  
and stability of skyrmions with the topological charge 2. Dynamics of skyrmions in frustrated magnets is different from that in chiral magnets because of the additional collective degree of freedom - skyrmion helicity. The coupling between the helicity and the skyrmion center-of-mass motion leads to interesting dynamical effects.

\section*{The model}

We consider classical spins, $\mathbf{S}_i$, of unit length  on a triangular lattice in the $xy$-plane with ferromagnetic nearest-neighbour (NN) and antiferromagnetic next-nearest-neighbour (NNN)  exchange interactions:
\begin{align}
E=
-J_1 \sum_{\langle i,j\rangle}\mathbf{S}_i\cdot\mathbf{S}_j&+J_2 \sum_{\langle\langle i,j\rangle\rangle}\mathbf{S}_i\cdot\mathbf{S}_j
\nonumber\\
&-h \sum_iS_i^z-\frac{K}{2}\sum_i(S_i^z)^2.
\label{energy}
\end{align}
where $\langle i,j \rangle$ and $\langle\langle i,j\rangle\rangle$ denote pairs of NN and NNN spins, respectively, and $J_1,J_2>0$. The third and the fourth terms describe, respectively, the interaction with the magnetic field parallel to the $z$ axis and the single-ion magnetic anisotropy.

The critical ratio $J_2/J_1$ for appearance of magnetic spirals and other modulated spin states is $1/3$. This can be understood by considering a modulation with the wave vector along the $y$ axis parallel to a NNN bond (see inset in Fig. \ref{fig:states}b), in which case the  model effectively reduces to a spin chain running in the $y$ direction with the ferromagnetic NN interaction, $J'_1 = 2(J_1 - J_2)$, and the antiferromagnetic NNN interaction,  $J'_2 = J_2$. A spiral state in a chain appears for $J'_2/J'_1> \frac{1}{4}$, 
which corresponds to $J_2/J_1 > \frac{1}{3}$. 
In what follows, the value of $J_1$ is set to 1.

 
\section*{Phase diagram} 

\begin{figure}
\includegraphics[width=8.7cm]{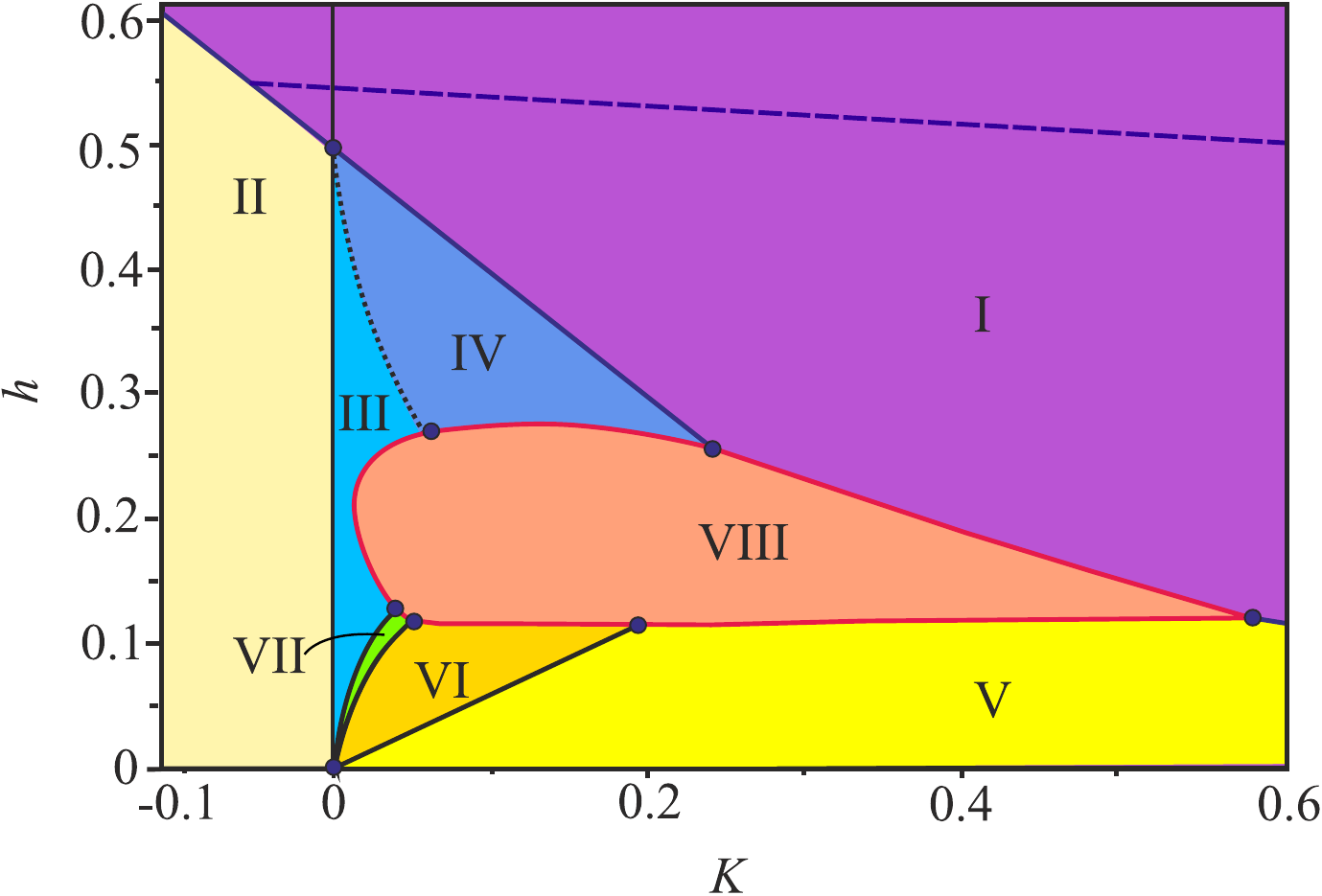}

\caption{\label{fig:phasediagram}  \textbf{Zero temperature phase diagram of the frustrated triangular antiferromagnet.} The eight phases are: (I) fully polarized ferromagnetic state, (II) conical spiral, (III) 2\textit{q}-state, (IV) 2\textit{q}$^{\prime}$-state, (V) vertical spiral, (VI)  vertical spiral plus two in-plane sinusoidal modulations, (VII) flop state and (VIII) skyrmion crystal. Dashed line is the upper critical field, above which isolated skyrmions are unstable. The NNN exchange constant, $J_2 = 0.5$, magnetic anisotropy, $K$, and magnetic field, $h$, are measured in units of the NN exchange constant, $J_1 = 1$. }
\end{figure}

\begin{figure}
\includegraphics[width=8.7cm]{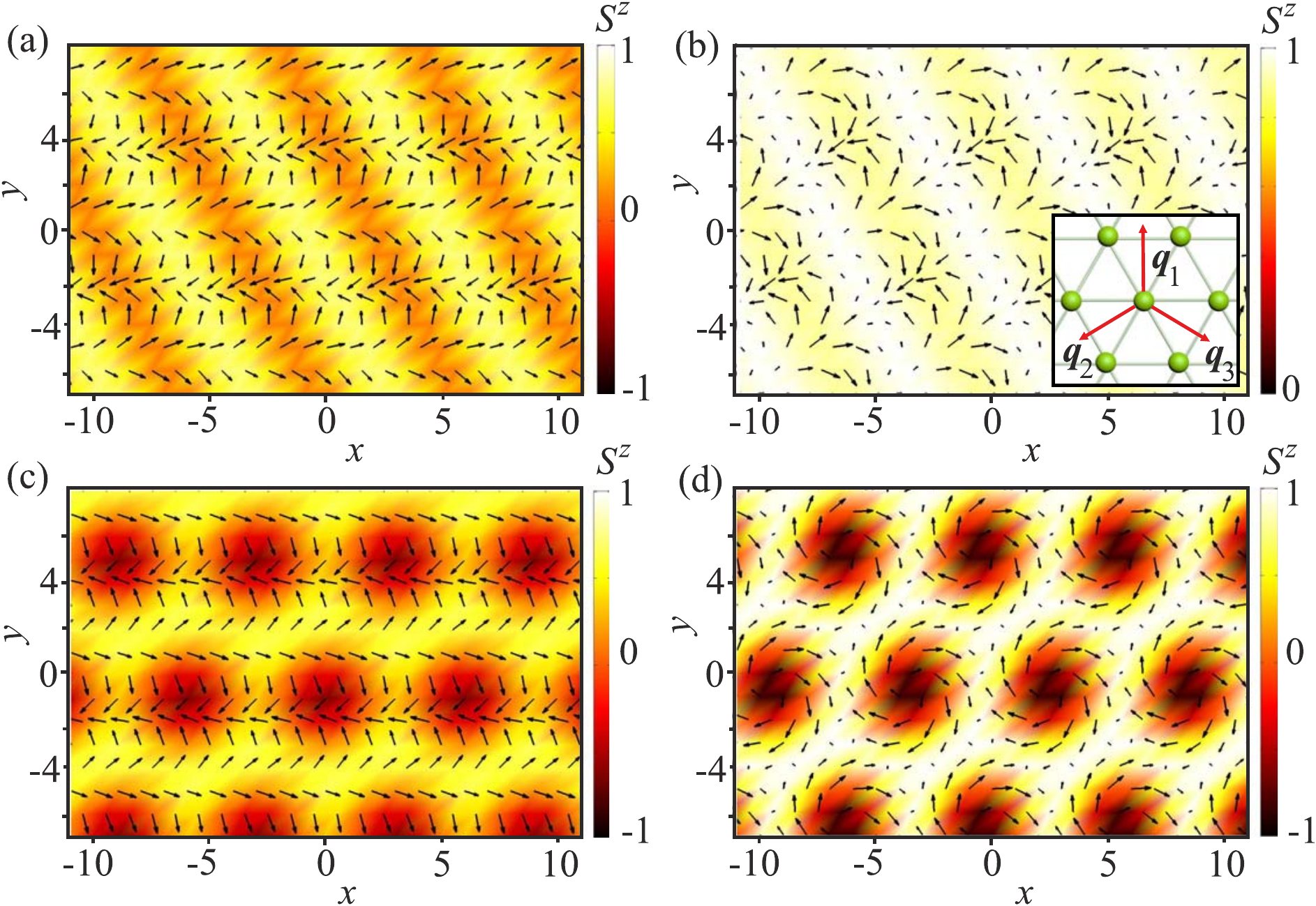}

\caption{\label{fig:states} \textbf{Multi-$q$ states.}  In-plane components (arrows) and out-of-plane components (color) of spins in four  selected states from the phase diagram Fig.~\ref{fig:phasediagram}. (a) 2\textit{q}-state (state III), (b)  2\textit{q}$^\prime$-state (state IV), (c) Flop state (state IV) and (d) Skyrmion crystal (state VIII). The inset shows the wave vectors of the three fundamental  modulations in the multi-$q$ states. 
} 
\end{figure}

Figure~\ref{fig:phasediagram} shows the zero-temperature phase diagram of the model in the $(K,h)$-plane for $J_2 = 0.5$, which counts eight phases. Their spin configurations can be described in terms of the Fourier series,
\begin{equation}
\bm S_i = \bm A_0 + \frac{1}{2} \sum_{\alpha=1}^3 \left(\bm A_\alpha e^{i\bm q_\alpha \cdot \bm x_i} + c.c.\right)+\mbox{higher harmonics}, 
\end{equation}
where $\bm q_1 = q(0,1)$, $\bm q_2 = q(-\frac{\sqrt{3}}{2},-\frac{1}{2})$ and $\bm q_3 = q(+\frac{\sqrt{3}}{2},-\frac{1}{2})$ are the wave vectors of the three fundamental modulations minimizing the exchange energy (inset in Fig. \ref{fig:states}b), such that $\bm q_{1}+\bm q_2+\bm q_3 = 0$. In all states, except the state VII, the uniform spin component $\bm A_0$ is parallel to the applied magnetic field: $\bm A_0 = (0,0,A_0^z)$. The amplitudes of higher harmonics, i.e. modulations with other wave vectors from the triangular lattice in the reciprocal space spanned on $\bm q_1$ and $\bm q_2$, are relatively small. The eight ground states of the model are:
\newline
(I)  The fully polarized ferromagnetic state with $A_0^z = 1$ and $\bm A_\alpha = 0$.
\newline
(II) The conical spiral state, which occupies the region of $K < 0$ (easy plane anisotropy) and $h < h_\ast(K) = \frac{(3J_2-1)^2}{J_2}-K$. This is a single-$q$ state with only one non-zero $\bm A_\alpha$ describing a circular spiral in the horizontal plane, e.g., $A_1^y = \pm i A_1^x$ and $A_1^z = 0$, $\pm$ sign corresponding to two opposite directions of spin rotation in the spiral. 
\newline
(III) The 2\textit{q}-state with $|\bm A_1| > |\bm A_2| \gg |\bm A_3|$ (see Fig.~\ref{fig:states}a). For $K>0$ (easy axis anisotropy) the conical state acquires the second spiral modulation in the horizontal plane, e.g., with the wave vector $\bm q_2$ and the opposite sense of spin rotation compared to that in the first spiral: $A_2^y = \mp i A_2^x, A_2^z = 0$. This `2\textit{q}-state' has also a small sinusoidal spin modulation parallel to the $z$ axis with the wave vector $\bm q_3$: $\bm A_3 = (0,0,A_3^z)$. As explained in the next section, the sinusoidal modulation  stabilizes the `2\textit{q}-state' state. For small positive $K$, $A_2 \propto \sqrt{K}$, i.e. $K = 0$ is a critical line.
\newline 
(IV) The 2\textit{q}$^{\prime}$-state, which is the same as the state III but with $|\bm A_1| = |\bm A_2|$ (see Fig.~\ref{fig:states}b). As $K$ increases, the amplitude of the second spiral grows until it becomes equal to that of the first spiral. This transition is marked by dotted line in Fig.~\ref{fig:phasediagram}.  
\newline
(V) The spiral in a vertical plane, e.g., $\bm A_1 = (i\cos(\chi)A_{\parallel},i\sin(\chi)A_{\parallel},A_{\perp})$ and $A_\parallel \approx A_\perp$, where the angle $\chi$ describes the rotation of the spiral plane around the $z$ axis.
This is the ground state for low applied fields and $K>0$. The vertical spiral is not perfectly circular and has higher harmonics.
\newline
(VI) The spiral in a vertical plane with two sinusoidal modulations in the direction normal to the spiral plane, e.g., the spiral in the $yz$ plane with the wave vector $\bm q_1$ and $\bm A_1 \approx   A_1(0,i,1)$ and the sinusoidal modulations with the wave vectors $\bm q_2$ and $\bm q_3$ along the $x$ direction: $\bm A_2 = (A_2,0,0)$ and $\bm A_3 = (-A_2^{\ast},0,0)$. The sinusoidal modulations appear above a critical magnetic field, $h_1(K)$, and the transition between the states V and VI is of second order. 
\newline
(VII) The `flop state' (Fig.~\ref{fig:states}c). As magnetic field increases further, the spiral plane changes from vertical to horizontal. This flop transition is not instantaneous, but occurs in a narrow field interval, 
$h_2(K)<h<h_3(K)$, in which  the spiral plane as well as the direction of the sinusoidal modulations rotate. 
The orientation of the spiral plane changes abruptly at $h_2$ and at $h_3(K)$ the system undergoes a first order transition to the 2\textit{q}-state. The flop state is the only state, in which the uniform magnetization $\bm A_0$ has a non-zero component parallel to the $xy$ plane.
\newline               
(VIII) The triangular skyrmion (or antiskyrmion) crystal state (Fig.~\ref{fig:states}d) with equal amplitudes of the  three fundamental modulations: 
\begin{equation}\label{eq:Skx}
\bm A_\alpha = Ae^{i \phi_{\alpha}}(-i\sin(\chi_\alpha),i\cos(\chi_\alpha),1).
\end{equation}
These modulations are spirals in three vertical planes rotated with respect to each other by $\pm 120^{\circ}$: $\chi_{\alpha}  = \chi + v (\alpha - 1) \frac{2\pi}{3}$ ($\alpha = 1,2,3$), where $v = \pm 1$ is the skyrmion vorticity describing the sense of rotation of the in-plane components of spins along a contour encircling the skyrmion centre \cite{Nagaosa2013}. The sum of the phases $\phi_1+\phi_2+\phi_3$ is either $0$ or $\pi$ and the $z$-component of spin in the centre of each skyrmion is $S^z=\cos(\phi_1+\phi_2+\phi_3)$. The topological charge, $Q$, of skyrmions forming the crystal is given by $Q = v \cos(\phi_1+\phi_2+\phi_3)$. The spin direction in the skyrmion centre is opposite to that of the applied magnetic field. Because of the arbitrary sign of vorticity, the topological charge of skyrmions can be both $+1$ and $-1$ for any direction of the field. This degree of freedom does not exist for skyrmions in chiral magnets and magnetic bubbles \cite{Okubo2012}. The SkX occupies a significant part of the phase diagram and all its boundaries are lines of first-order transitions.


\section*{Instabilities of spiral states}

To clarify the origin of multi-\textit{q} states, we discuss a continuum model of the frustrated triangular magnet, applicable when the modulation period is much larger than the lattice constant ($q \ll 1$):
\begin{equation}
E = \int\!\!d^2x \left( \frac{1}{2} \bm S \hat{D} \bm S + \frac{K}{2} (S^z)^2 - h S^z \right),
\label{eq:continuum}
\end{equation}
where the first term is the exchange energy, the second term is magnetic anisotropy and the last term is the Zeeman energy. 
The operator $\hat{D}$ in the momentum representation can be expanded in powers of the wave vector $\bm q$:
 \begin{align}
D_{\bm q} &= 
d_0 + d_2 q^2 + d_4 q^4 + d_6 q^6\nonumber\\
&+ d'_6 \left(q_x^6 -15 q_x^4 q_y^2+15 q_x^2 q_y^4-q_y^6\right)+\ldots.
\end{align}
The anisotropy in the reciprocal space first appears in the sixth order of the expansion, resulting in 6 minima of $D_{\bm q}$ at $\bm q = \pm \bm q_1, \pm \bm q_2, \pm \bm q_3$. It is convenient to add a constant to energy to make $D_{\bm q_{1}} = 0$. 

The six-fold degeneracy of the exchange energy minimum is the source of complexity. 
Another important factor is the relative rigidity of single-spiral states, which for $K,h \neq 0$ makes the states with several coexisting modulations, spiral or sinusoidal, energetically more favourable, even though the multi-\textit{q} states have a larger exchange energy due to higher harmonics induced by the local constraint $\bm S^2(\bm x) = 1$.  
%

To prove this point, we consider the case of weak anisotropy, $|K| \ll 1$, and magnetic fields just below $h_{\ast}(K)=  D_0 - K$, at which the transition to the saturated ferromagnetic state occurs. 
Then the in-plane component, $\bm m$, of spin   
$  
\bm S = \bm m + \sqrt{1-m^2}\hat{\bm z}
$
is small and the energy (\ref{eq:continuum}) can be expanded in powers of $\bm m$:
\begin{align}
E &\approx \int\!\!d^2x \left[ \varepsilon_{\rm FM}+\frac{1}{2} \bm m (\hat{D}+\delta h)\bm m\right.\nonumber\\
 &+\left.\frac{1}{8} m^2(\hat{D}+\delta h-K)m^2 \right],
\end{align}
where $\varepsilon_{\rm FM}$
is the energy density of the saturated ferromagnetic state and $\delta h = h - h_{\ast}(K)<0$. 
For $K<0$,  the ground state of the model is the conical spiral (state II): $\bm m_0 = A\left(\hat{\bm x}\cos (\bm q_1\!\cdot\!\bm x + \alpha) \pm \hat{\bm y}\sin( \bm q_1\!\cdot\!\bm x + \alpha)\right)$ with $A = \sqrt{\frac{2|\delta h|}{h_{\ast}(K)}}$ and an arbitrary phase $\alpha$. 
For $K>0$, the conical spiral state becomes unstable against formation of an additional spin modulation. A small deviation from the conical spiral, $\bm k = \bm m - \bm m_0$, changes energy by
$
\delta E \approx \frac{1}{2}\int\!\!d^2x\left[\bm k \hat{D} \bm k + (\bm k\!\cdot\!\bm m_0)(\hat{D} - K)(\bm k\!\cdot\!\bm m_0)\right].
$
An additional modulation satisfying
\begin{equation}
\left\{
\begin{array}{ccc}
\hat{D} \bm k &=& 0,\\
\hat{D} (\bm k\!\cdot\!\bm m_0)&=& 0,
\end{array}
\right.
\label{eq:conditions1}
\end{equation}
changes energy by $\delta E = -\frac{K}{2}\int\!\!d^2x(\bm k\!\cdot\!\bm m_0)^2<0$, for $K>0$. The solution of Eqs. (\ref{eq:conditions1}) is the second $xy$-spiral with the wave vector $\bm q_2$ (or $\bm q_3$) and the opposite direction of spin rotation: $\bm k = B\left(\hat{\bm x}\cos (\bm q_2\!\cdot\!\bm x + \beta) \mp \hat{\bm y}\sin( \bm q_2\!\cdot\!\bm x + \beta)\right)$. The two-spiral state III has a lower energy because of the
small sinusoidal modulation of the $z$-component of spin with the wave vector $\bm q_3$, $\delta S^z =  -(\bm k \cdot \bm m_0) = -AB \cos (\bm q_3 \cdot \bm x - \alpha -\beta)$, necessary to preserve the spin length. This sinusoidal modulation also has the minimal exchange energy, but its anisotropy energy for $K>0$ is lower than that of the $xy$ spirals, which makes the conical state unstable. 

Next we discuss the instability of the vertical spiral (state V),  
\begin{equation}
\bm S = \pm \hat{\bm y} \sin \phi + \hat{\bm z} \cos\phi,
\label{eq:yzspiral}
\end{equation}
for $h,K \ll 1$.
The magnetic field and easy axis parallel to the spiral plane deform the spiral: $\phi \approx \bm q_1  \cdot \bm x - a \sin (\bm q_1 \cdot \bm x) - b \sin \left(2 \bm q_1 \cdot \bm x\right)$ with $a \approx \frac{2h}{D_{0}+D_{2\bm q_1} }$ and $b \approx \frac{K}{D_{3\bm q_1}}$, corresponding to 
\begin{equation} 
S^z \approx \frac{a}{2} + \cos (\bm q_1 \cdot \bm x) - \frac{a}{2} \cos \left( 2\bm q_1 \!\cdot\! \bm x \right) - 
\frac{b}{2} \cos \left( 3\bm q_1 \!\cdot\! \bm x\right).
\end{equation}
Importantly, the uniform magnetization equals the amplitude of the energetically costly second harmonic. This makes the  
magnetic susceptibility of the $yz$-spiral low and leads to 
appearance of an $x$-component of spin above a critical magnetic field.  Expanding the spin energy in powers of $S^x$ around Eq. (\ref{eq:yzspiral}), we obtain  
$
\sin (\phi) \hat{D} S^x - S^x \hat{D} \sin (\phi) = 0,
$
the solution of which is the sum of two sinusoidal spin modulations with the wave vectors $\bm q_2$ and $\bm q_3$ plus relatively small higher harmonics:
$
S^x \approx  C \left[\cos (\bm q_2 \cdot \bm x + \gamma) - \cos (\bm q_3 \cdot \bm x - \gamma)\right],
$
$\gamma$ being an arbitrary phase. The second-order transition to the non-coplanar 3\textit{q}-state VI occurs at $h_{1}(K) \approx K \frac{(D_0 +D_{2\bm q_1})}{2 D_{2\bm q_1}}$, when the lowering of the Zeeman energy due to the larger magnetization of the 3\textit{q}-state compensates the increase of the anisotropy energy. 

Similarly, the SkX owns its stability partly to the low anisotropy energy (as it is composed of three vertical spirals) and partly to its large magnetic moment. The latter is the consequence of the constraint $\bm S^2 = 1$, which is impossible to maintain without adding to the $3q$-state a large uniform magnetization.

The discussion of instabilities of spiral states in the continuum model of the frustrated triangular magnet also shows that the phase diagram Fig.~\ref{fig:phasediagram} is generic, a minor difference being that in the discrete model with $J_2/J_1 =1/2$ the wave vectors of the three fundamental modulations are commensurate with the lattice and constant throughout the phase diagram, while in the continuum model $q$ is a function of $K$ and $h$.  

 
\section*{Helicity reversals and interaction between skyrmions}

\begin{figure}
\includegraphics[width=8.7cm]{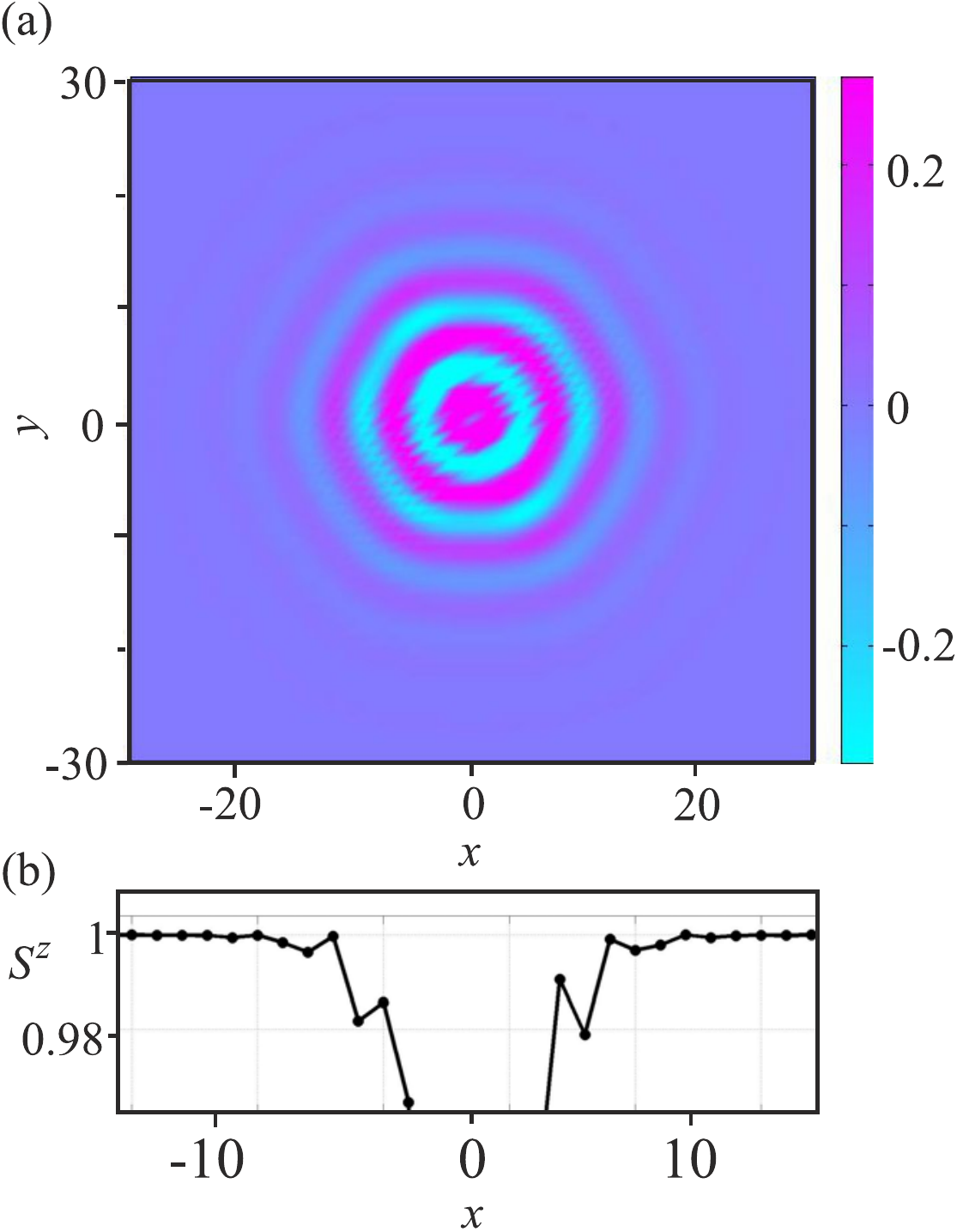}
\caption{\label{fig:fanoscillations} \textbf{Helicity reversals.}  (a) False colour plot of the toroidal moment density in the skyrmion with the helicity angle $\chi = \pi/2$. (b) Fan-like oscillations of the out-of-plane spin component  $S^z$. The skyrmion centre is located at $x = 0$. } 
\end{figure}

Next we discuss properties of isolated skyrmions. The asymptotic of the in-plane spin components at a large distance $r$ from the skyrmion centre is given by 
\begin{equation}
\left(
\begin{array}{c}
S^x\\
S^y
\end{array} 
\right)
\approx u(r)
\left( 
\begin{array}{cc}
\cos \varphi & - v \sin \varphi\\
v \sin \varphi & \cos \varphi
\end{array}
\right)
\left( 
\begin{array}{c}
\cos \chi\\
\sin \chi
\end{array}
\right),
\label{eq:asymptotics1}
\end{equation}
where $\varphi$ is the azimuthal angle in the $xy$ plane and $v$ and $\chi$ are, respectively, the skyrmion vorticity and helicity angle introduced in Eq. (\ref{eq:Skx}) (for simplicity, we consider large skyrmions and replace the discrete rotational symmetry of the triangular lattice by the continuous one). 
In absence of in-plane magnetic anisotropy, the helicity angle $\chi$ is arbitrary, $\chi = 0,\pi$ corresponding to skyrmions carrying a monopole moment,  $A \propto \sum_j \bm r_j \cdot \bm S_j$, and $\chi = \pm \pi/2$ corresponding to skyrmions with a toroidal moment, $T^z \propto \sum_j [\bm r \times \bm S_j]^z$ \cite{Spaldin2008}. Arbitrary $\chi$ is a zero mode, which skyrmions in chiral magnets and magnetic bubbles do not have.

In addition, the toroidal and monopole moment densities show periodic sign reversals as the distance $r$ from the center increases (see Fig.~\ref{fig:fanoscillations}), similar to the helicity reversals observed in magnetic bubbles in a ferrimagnetic hexaferrite, where $\chi$ can be both $+\pi/2$ and $-\pi/2$ \cite{Yu2012}. We find it more convenient to keep the helicity angle fixed and associate these oscillations with the sign changes of $u(r)$ in Eq. (\ref{eq:asymptotics1}), 
\begin{equation}
u(r) \sim \frac{A}{\sqrt{r}}e^{-\sqrt{\frac{\kappa^2-q^2}{2}}r}\cos \left(\sqrt{\frac{\kappa^2+q^2}{2}}r + \psi\right),
\label{eq:assymptotics2}
\end{equation}
where $q \approx \sqrt{\frac{d_2}{2d_4}}$ is the length of the minimal-energy wave vector and $\kappa^4 = \frac{K+h}{d_4}$. The sign reversals of $u(r)$ correspond to fan-like oscillations of spins around the $z$ axis, whose amplitude decreases exponentially with $r$ for $\kappa > q$ or, equivalently, $h > h_\ast(K)$. This condition ensures that the skyrmion energy is finite and, hence, the lower critical field for  stability of isolated skyrmions coincides with the line separating the fully polarized state from the conical spiral state, for $K<0$, and the 2\textit{q}$^\prime$-state, for $K>0$. There is also an upper critical field (dashed line in Fig.~\ref{fig:phasediagram}), above which skyrmions collapse.

\begin{figure}
\includegraphics[width=8.7cm]{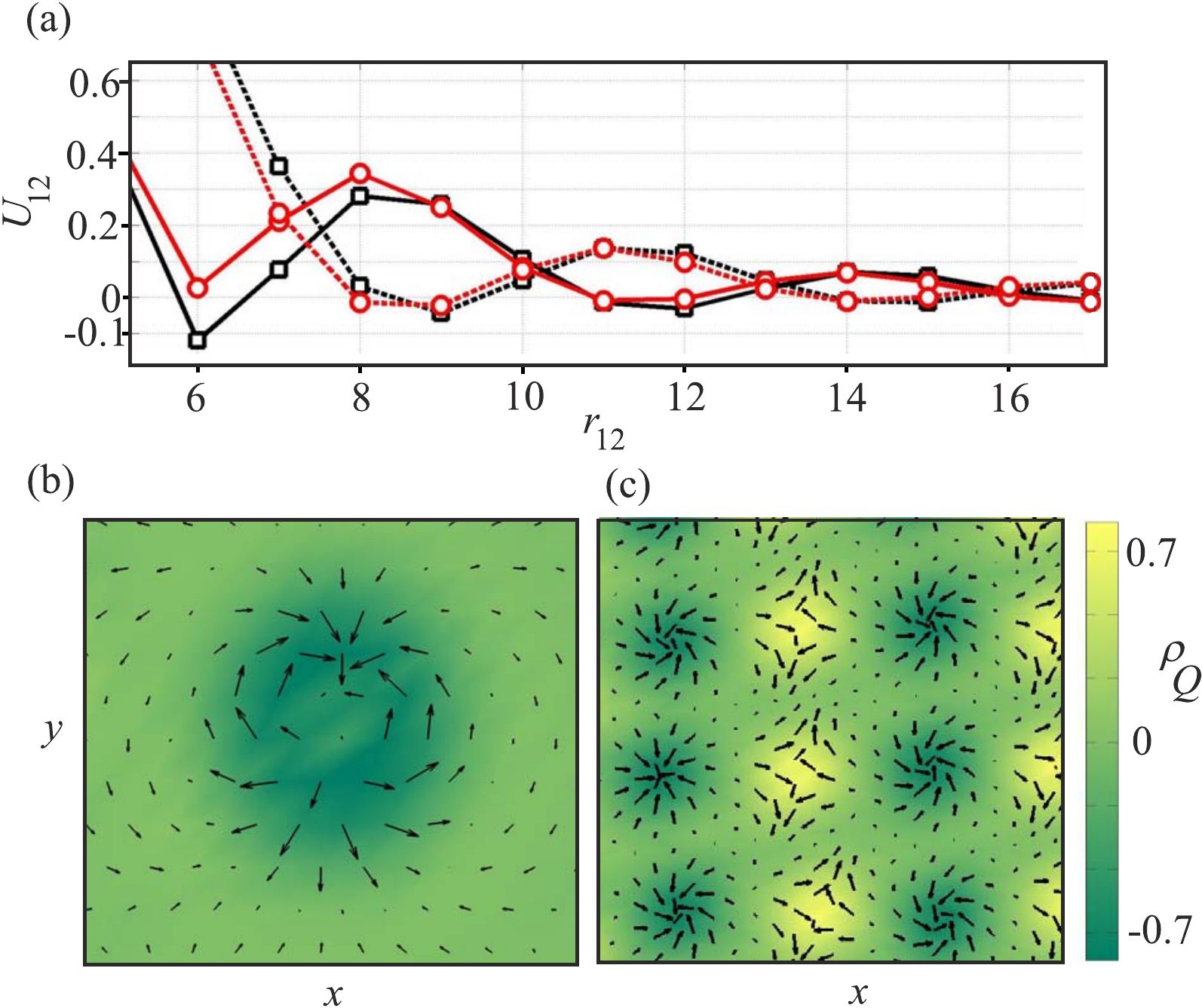}
\caption{\label{fig:interactions} \textbf{ Unusual features of skyrmions in frustrated magnets.} (a) The interaction energy, $U_{12}$, vs the distance $r_{12}$ between two skyrmions (black lines with rectangular markers), and skyrmion and antiskyrmion (red lines with circular markers). Solid (dashed) lines show the interaction for equal (opposite) helicities of the topological defects. Skyrmion positions and helicities were fixed by the condition $S_z=-1$ at the skyrmion centre and $\theta_i = \theta$ at the six nearest-neighbour sites.  
(b) Metastable skyrmion with topological charge $Q=2$.  (c) Metastable skyrmion-antiskyrmion crystal with a rectangular lattice. In panels (b) and (c) color indicates the topological charge density,  $\rho_{Q}$, while arrows show the in-plane components of spins.}
\end{figure}

The fan oscillations of spins in skyrmion give rise to sign changes of the skyrmion-skyrmion interaction $U(r_{12})$ (Fig.~\ref{fig:interactions}a).
This interaction depends on the topological charges of skyrmions. For brevity, we call a skyrmion with positive vorticity and $Q = -1$  a skyrmion, while a skyrmion with $v = -1$ and $Q = +1$ is called an anti-skyrmion. Two skyrmions or two antiskyrmions attract each other at distances of the order of the skyrmion diameter, the maximal reduction of energy being $\sim 10 \%$ of the skyrmion energy. Because of the attraction, the skyrmion crystal phase extends to $h > h_{\ast}(K)$, as the energy per skyrmion in the crystal can be negative even when the energy of an isolated skyrmion is positive. Isolated skyrmions tend to form clusters with a short-range crystal order, resembling the clustering of vortices in 1.5-type superconductors \cite{Babaev2012}.  The oscillating interactions between skyrmions in the frustrated magnet are in sharp contrast with skyrmion-skyrmion interactions in non-centrosymmetric magnets, which do not show fan-like oscillations and repel each other at all distances \cite{Leonov11}.   

Interactions between skyrmions also depend on their   helicities:  $U_{12}(r) \approx  V(r) + W(r)\cos(\chi_1 - \chi_2)$ (see Fig.~\ref{fig:interactions}a), which couples the helicity dynamics to the translational motion of skyrmions.
Another difference from chiral magnets is the existence of the locally stable  skyrmion with topological charge $Q = \pm 2$, shown in  Fig.~\ref{fig:interactions}b, which can be considered as a tightly bound state of two skyrmions with opposite helicities, similar to the magnetic bubble with $Q = 2$ observed in thin films of a CMR manganite \cite{Yu2014}. The energy of the $Q=2$ skyrmion is, however, larger than the energy of two $Q = 1$ skyrmions.

The helicity dependence of the skyrmion-antiskyrmion interaction is more complex: it is approximately proportional to $\cos(2 \alpha_{sa} +\chi_s - \chi_a)$, where $\alpha_{sa}$ is the angle between the line connecting skyrmion with antiskyrmion and the $x$ axis, with respect to which the skyrmion and anstiskyrmion helicity angles, $\chi_s$ and $\chi_a$,  are defined. 
Figure~\ref{fig:interactions}a shows skyrmion-antiskyrmion interactions for $\chi_s = \chi_a$ and $\alpha_{sa} = 0$. The minimum of the skyrmion-antiskyrmion potential occurs at larger distances and is more shallow than that of the skyrmion-skyrmion potential. One can form a metastable rectangular crystal with alternating columns of skyrmions and antiskyrmions (Fig.~\ref{fig:interactions}c). Such a crystal with zero topological charge has a higher energy than a purely skyrmion (or purely antiskyrmion) crystal, which is important for observation of Topological Hall Effects in frustrated magnets.




\section*{Magnetoelectric coupling}

All modulated states in the phase diagram Fig.~\ref{fig:phasediagram} break inversion symmetry and can induce an electric polarization, $\bm P$, in a magnetic insulator. 
The two possible directions of spin rotation in spirals correspond to two opposite directions of $\bm P$ \cite{Cheong2007}.

Consider, for example, a quasi-two-dimensional magnet with the `eclipsed' triangular spin layers stacked on top of each other \cite{Cava2011} and the symmetries of the crystal lattice being inversion $I$, three-fold axis $3_z$ and the two-fold axis $2_x$ (e.g. space groups $P\bar{3}m_1$, $R\bar{3}m$). This allows for three independent Lifshitz invariants describing the magnetically-induced electric polarization, which in momentum representation have the form:
\begin{eqnarray}\label{eq:mecoupling}
&&H_{\rm ME} = - {\Im m} \sum_{\bm q}
\big[
g_1 E^z S^{z}_{-\bm q}(q^x S^{x}_{\bm q}+q^y S^{y}_{\bm q}) 
\nonumber\\ 
&&+
g_2 (q^x E^y-q^y E^{x})
S^{x}_{-\bm q}S^{y}_{\bm q}\\
&&
+g_3S^{z}_{-\bm q}\left((q^x E^x-q^y E^{y})
S^{y}_{\bm q}+(q^x E^y+q^y E^{x})
S^{x}_{\bm q}\right)
\big].
\nonumber
\end{eqnarray} 
The polarization, $\bm P = - \frac{\partial H_{ME}}{\partial \bm E}$, described by the first two terms lies in the spiral plane and is orthogonal to the spiral wave vector, as is the case for $\bm P$ induced by a cycloidal spiral through the inverse Dzyaloshinskii-Moriya mechanism  \cite{Katsura2005,Sergienko2006,Mostovoy2006}. The third term describes other symmetry-allowed directions of polarization, in particular, $\bm P$  parallel to the wave vector of a helical spiral  \cite{Arima2007}. 
 
\begin{figure}
\includegraphics[width=8.7cm]{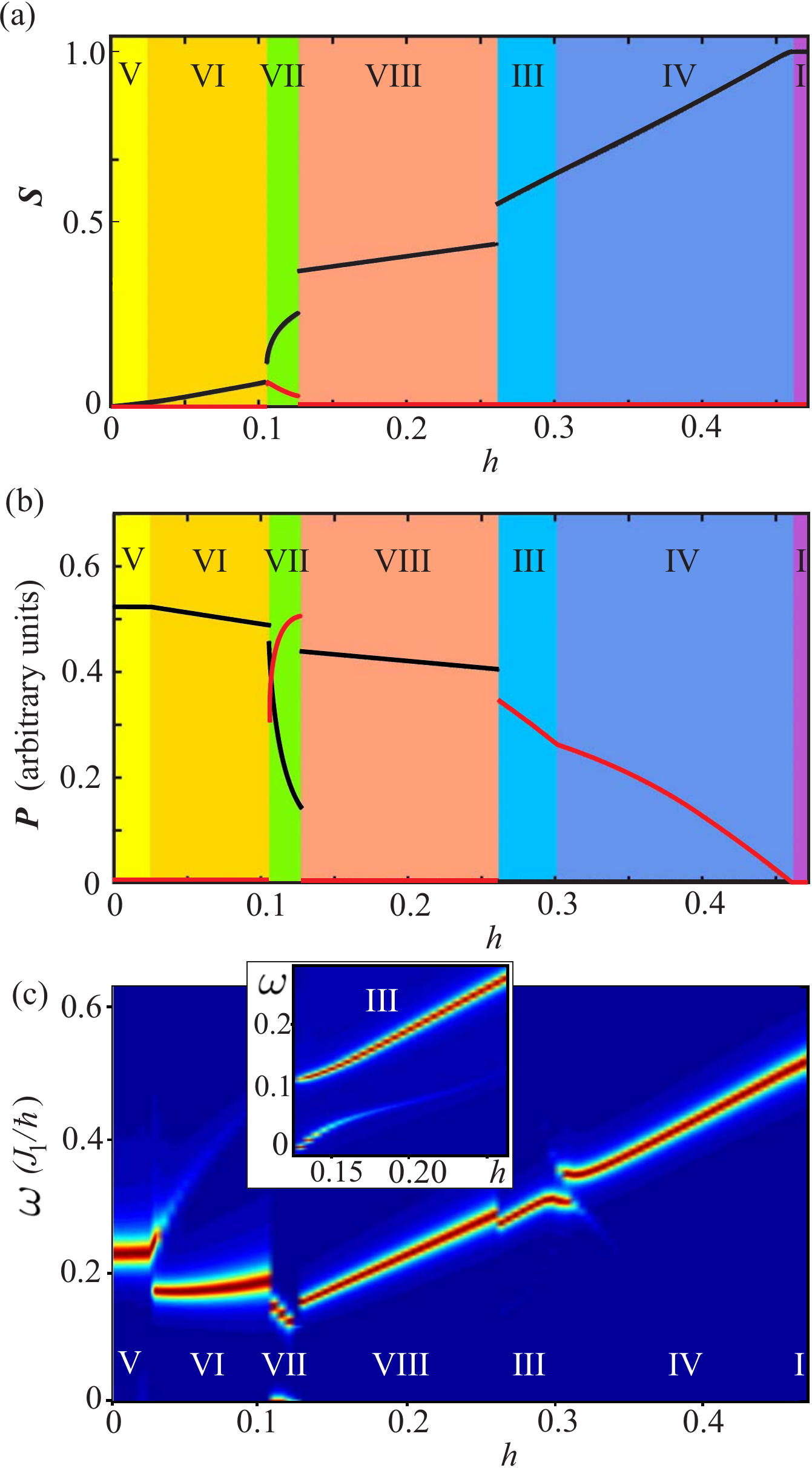}
\caption{
\label{fig:polarization} \textbf{ Multiferroic and dynamical properties of multi-$q$ states.} (a) Magnetic field dependence at $K = 0.04$ of (a) the average spin vector $\langle \bm S \rangle$,  (b) the electric polarization vector $\bm P$ and (c) the imaginary part of the in-plane magnetic susceptibility,  $\chi^{\prime\prime}(\omega)$ (false color plot). 
Black and red lines in panels (a) and (b) are the out-of-plane and in-plane components of the vectors. The electric polarization was calculated
using Eq. (\ref{eq:mecoupling}) for $g_1=g_2$ and $g_3=0$. The inset in panel (c) shows magnetic modes in the 2\textit{q}-state near the transition to the flop state (see text).
}
\end{figure}

Figure~\ref{fig:polarization}b shows magnetic field dependence of the electric polarization in arbitrary units,   for $K = 0.04$, $g_2 = g_1$ and $g_3 = 0$. The states with spirals in a vertical plane, including the SkX, induce polarization in the $z$-direction (black line), while the states with spirals in the horizontal plane (conical spiral and the two-spiral states) induce an in-plane polarization (red line). In the flop phase the spiral plane rotates in an applied magnetic field and so does $\bm P$. The spontaneous in-plane magnetization, present in this phase (see Fig.~\ref{fig:polarization}a) allows for effective control of the polarization direction by magnetic field and \textit{vice versa}. 

The orientations of spiral planes and skyrmion helicity in realistic materials will be affected by the sixth-order in-plane magnetic anisotropy, which we neglected. The interplay between this weak anisotropy and the interaction of the electric polarization induced by the spiral with an applied electric field can lead to giant magnetoelectric effects, such as the electrically-induced rotation of the skyrmion helicity.    

\section*{Magnetic resonance}

We also studied collective spin dynamics for the states in the phase diagram Fig.~\ref{fig:phasediagram} induced by the ac magnetic field, $h_{\omega} \sin(\omega t)$, parallel to the $xy$ plane. 
Figure \ref{fig:polarization}c shows the false color plot of the imaginary part of the in-plane dynamical magnetic susceptibility  calculated along the $K=0.04$ line in the phase diagram (as in panels (a) and (b)) by solving Landau-Lifshitz-Gilbert equation (see Methods).  
The phase transitions between different states are reflected in sudden changes in the magnetic resonance spectrum.

\begin{figure}
\includegraphics[width=8.7cm]{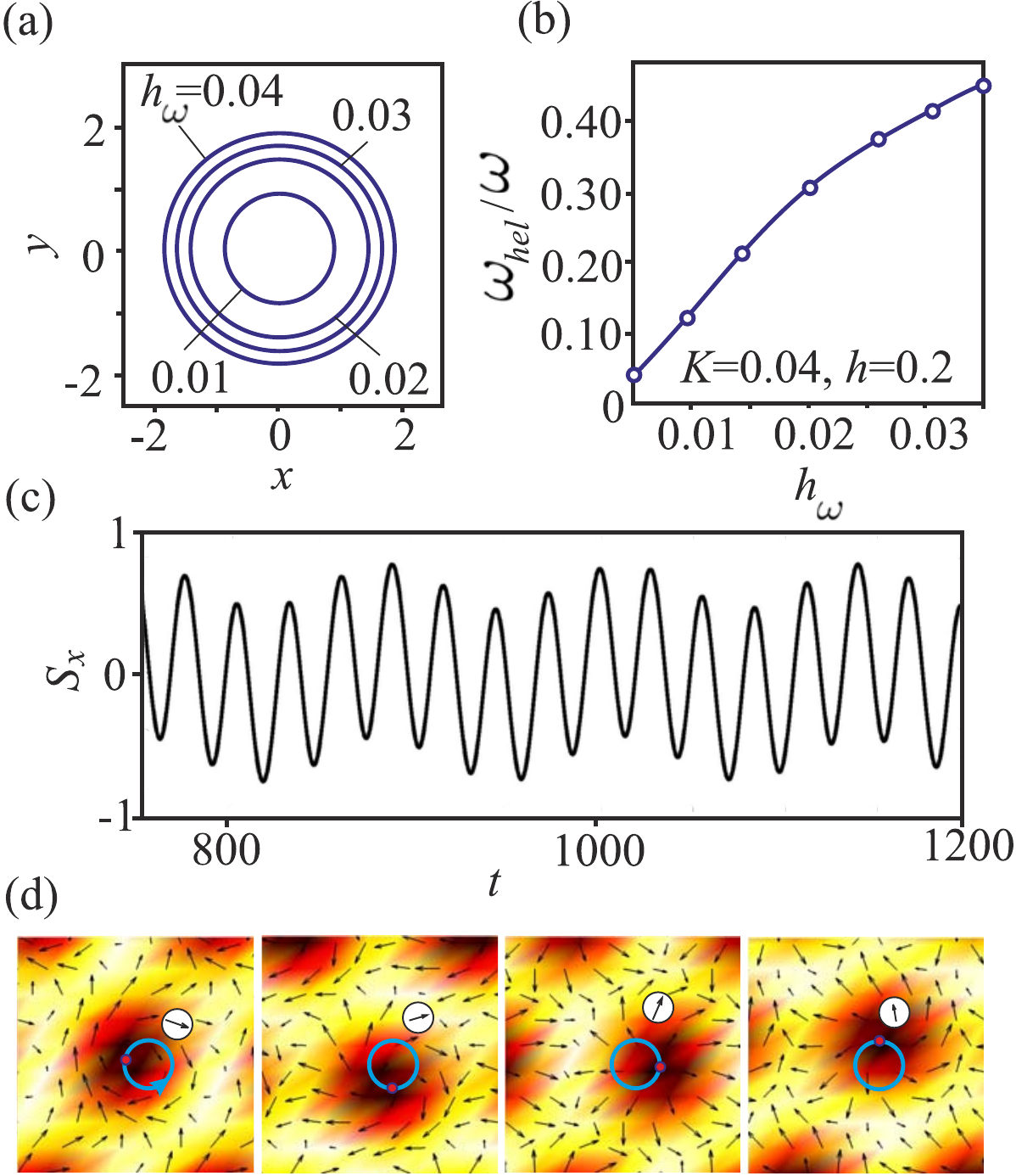}
\caption{
\label{fig:rotation}  \textbf{Coupled dynamics of skyrmion helicity and centre-of-mass.} (a) Circular trajectories spanned by the centres of rotating skyrmions in the in-plane ac magnetic field for various amplitudes of the ac magnetic field  $h_{\omega}$. (b)  Frequency of the helicity  rotation, $\omega_{hel}$, measured in units of the frequency of the ac field $\omega=0.224 J_1 \hbar^{-1}$ vs $h_{\omega}$. (c) Time evolution of the $x$-component of spin at a chosen site, marked by a circle in the series of snapshots (panel d), which also shows current positions of the skyrmion centre (red point) and its trajectory for $\textit{h}_{\omega}=0.01$. This calculation was performed for $h=0.2$ and $K=0.04$.
}
\end{figure}

While the SkX in chiral magnets shows two  resonances: one with the counterclockwise and another with the anti-counterclockwise rotation of the skyrmion centres \cite{Mochizuki12}, in the frustrated magnet we find only one mode, in which skyrmions rotate counterclockwise. 
Figure \ref{fig:rotation}a shows circular trajectories traced by the skyrmion centres for various amplitudes of the ac field and the resonant frequency $\omega=0.224 J_1 \hbar^{-1}$.

The field-induced spin dynamics in frustrated magnets is, however, more complex than in chiral magnets, because of the additional zero mode - helicity.  
This can be seen from the fact that the time dependence of an in-plane component of spin at a chosen site  (see  Fig.~\ref{fig:rotation}c) is a superposition of periodic oscillations with two different periods.
The shorter period corresponds to the rotation of the skyrmion centre, while the longer one results from a monotonous increase of the helicity angle, as evidenced by the series of snapshots shown in Fig. \ref{fig:rotation}d (see also the movie in Supplementary Material).   
The helicity dynamics is induced by the rotation of skyrmion centres through a non-linear coupling between the two modes, as can be seen from the non-linear dependence of the frequency of helicity rotations, $\omega_{hel}$, on the amplitude of the ac magnetic field (Fig. \ref{fig:rotation}b). The coupling between the helicity and translational dynamics must have important consequences for the current-induced motion of skyrmions in confined nanostructures.

Another remarkable feature of the magnetic resonance spectrum is the presence of the zero-frequency peak in the flop phase (Fig. \ref{fig:polarization}c) related to the non-zero spontaneous in-plane magnetization, which in absence of the in-plane magnetic anisotropy has an arbitrary direction. 
Surprisingly, a low-energy mode is also present in the 2\textit{q}-state near the first-order transition to the flop phase, as shown in the inset to Fig. \ref{fig:polarization}c (this transition does not actually occur at $K = 0.04$ because of the intervening SkX phase, which was artificially removed to obtain the susceptibility shown in the inset).

\section*{Conclusions}

In conclusion, we studied a new class of skyrmion materials -- frustrated magnets -- and clarified the mechanism for stabilization of various multi-\textit{q} states which does not involve thermal spin fluctuations. The crucial role in this mechanism is played by magnetic anisotropy, which together with an applied magnetic field leads to instability of spiral states. An interesting question is why the 2\textit{q} and skyrmion crystal states are favoured by both the entropic mechanism in absence of anisotropy \cite{Okubo2012} and the magnetic anisotropy at zero temperature? The common origin seems to be the soft magnetic excitation spectrum of  multi-\textit{q} states, which on the one hand lowers free energy of these states by increasing their entropy and, on the other hand, lowers their energy by enhancing susceptibility to magnetic field and anisotropy. Soft magnetic modes of multi-\textit{q} states, which can be excited by both electric and magnetic fields, lead to large magnetoelectric responses. 

Finally, we list a few requirements for stabilization of skyrmions in frustrated magnets following from our study, which can be used as guidelines in the search for new materials hosting skyrmions: (1) spiral spin ordering in zero magnetic field resulting from competing exchange  interactions; (2) three-fold or six-fold symmetry axis, which ensures that the spirals with the three wave vectors, $\bm q_1$, $\bm q_2$ and $\bm q_3$ forming the skyrmion crystal, have the same energy; although cubic lattice symmetry does not favour the three vectors lying in the same plane, it results in a weak dependence of the exchange energy on $\bm q$, which may be sufficient;   
(3) easy axis magnetic anisotropy axis stabilizing multi-\textit{q} states. Although these requirements are not very demanding and some of them are met in NiGa$_2$S$_4$ \cite{Nakatsuji2005,Stock2010}, $\alpha$-NaFeO$_2$ \cite{McQueen2007}, dihalides NiBr$_2$ and Fe$_x$Ni$_{1-x}$Br$_2$ \cite{Day1976,Regnault1982,Moore1985}, it is a challenge to find a material which satisfies them all.

\section*{Methods}

The energy (\ref{energy}) has been minimized using the iterative simulated annealing procedure and a single-step Monte-Carlo dynamics with the Metropolis algorithm (for details see Ref.~\cite{Leonov11}). We imposed periodic boundary conditions and performed simulations for lattices of different sizes to check the stability of the numerical routine. 

The dynamical magnetic susceptibility for all multi-\textit{q} states and the time evolution of spins in the oscillating magnetic field have been obtained by solving the Landau-Lifshitz-Gilbert (LLG) equation using the fourth-order Runge-Kutta method. The dimensionless damping parameter $\alpha=0.01$. To obtain initial spin configurations, we used the annealing technique supplemented by a further LLG relaxation for $\textit{h}_{\omega}=0$. When the convergence of the spin configurations was reached, we switched on a periodic magnetic field in the plane $xy$ to study the spin dynamics (Fig.~\ref{fig:rotation}) or applied a magnetic field pulse at $t = 0$ and monitored the response, to obtain the magnetic susceptibility (Fig. ~\ref{fig:polarization}c).

The skyrmion-skyrmion and skyrmion-antiskyrmion interaction potentials shown in Fig.~\ref{fig:interactions}, were calculated by minimizing the spin energy with the constraint, $S^z = -1$, imposed at the centres of the topological defects. To calculate the helicity dependence of the interactions, we constrained the helicity angle at six sites neighbouring to the skyrmion centre and imposed $S^z = -0.427$ at these sites, which holds for an isolated skyrmion.

\textbf{Acknowledgements}

The authors would like to thank N. Nagaosa, 
Y. Tokura, 
U. K. R\"o{\ss}ler, 
A. N. Bogdanov, 
R. Wiesendanger  
and  
J. van den Brink 
for valuable discussions. This study was supported by the FOM grant 11PR2928.





%

\end{document}